# Live, Rich, and Composable: Qualities for Programming Beyond Static Text

Joshua Horowitz [1] and Jeffrey Heer [1]

[1] University of Washington, Seattle, WA

**Abstract**

Efforts to push programming beyond static textual code have sought to imbue programming with multiple distinct qualities. One long-acknowledged quality is *liveness*: providing programmers with in-depth feedback about a program's dynamic behavior as the program is edited. A second quality, long-explored but lacking a shared term of art, is *richness*: allowing programmers to edit programs though domain-specific representations and interactions rather than solely through text. In this paper, we map the relationship between these two qualities and survey past work that exemplifies them. We observe that systems combining liveness and richness often do so at the cost of an essential quality of traditional programming: *composability*. We argue that, by combining *liveness, richness, and composability*, programming systems can better capture the full potential of interactive computation without leaving behind the expressivity of traditional code. *Keywords*: Live programming, visual programming, end-user programming, composition, computational notebooks, GUIs.

## 1 Introduction

Alternatives to static textual code have been explored since the advent of interactive computing. The earliest user interfaces included systems for programmatic CAD [1] and flow-chart programming [2], [3]. These were soon followed by the spreadsheet, which remains an exemplar of end-user programming [4].

There are many reasons to depart from static text. The most prominent is the desire to make programming more accessible. Students of programming and potential end-user programmers face serious obstacles approaching conventional programming languages. Perhaps new interfaces could unlock programmatic power for more people. Others have focused instead on how static text forces all programmers (even experts) to interact with artifacts as diverse as data visualizations and interactive games through abstract, indirect representations. Perhaps new interfaces could bridge this gap, giving "creators… an immediate connection to what they create" [5].

Driven by these motives, researchers and tool-builders have built an enormous diversity of programming systems beyond static text, ranging from spreadsheets to block-based editors to exotic visual programming languages. It is clear that static text is a single point in vast space of possible programming environments. To make sense of what might otherwise be an overwhelming cloud of point examples, we will identify fundamental qualities that these designs advance.

We focus on two core qualities that have emerged in efforts beyond static text. The first quality is *liveness*: providing programmers with in-depth feedback about a program's dynamic behavior as the program is edited. This quality is well-established in the computer-science literature.[1] The second quality is as yet unnamed, though its pursuit is implicit in many projects: allowing programmers to work with domain-specific visualizations and interactions. We offer *richness* as a new term of art to better pin down this quality for analysis.

We look at these two qualities, surveying systems that exemplify them in different combinations. In the course of this analysis, we run into a third quality called *composability*: the ability to freely combine smaller programmed artifacts into larger ones, to accomplish larger goals. Unlike liveness and richness, this is not a quality static text lacks, which interactive programming systems strive to add to it. Rather, it is a familiar quality of static text which new programming systems must work hard to maintain.

We hypothesize that by combining *liveness*, *richness*, and *composability*, programming systems can meaningfully extend the capabilities of static text without losing its characteristic expressivity. While it is challenging to embody all of these qualities at once, several recent projects have made great strides towards this goal. We conclude with a discussion of these systems.

---

[1] Liveness was named as early as 1990 [6], and a workshop named after it has been running since 2013 [7].

## 2 Liveness

Programming is conventionally an opaque art. As a computer runs a program, it navigates through control structures, performing operations and constructing intermediate values. Programmers are tasked with carefully orchestrating this behavior, but oftentimes the behavior is invisible. No feedback is provided by the computer to the programmer to tell them what their program will do as they write it. Instead, programmers simulate in their heads how their program will run, without computational assistance. While debuggers can provide some of this information, they do so through a narrow peephole intended for extraordinary ("debugging") situations. While editing code, programmers generally work without visibility into the effects of their edits on run-time behavior.

*Live programming* refers to efforts to move away from this status quo, creating "programming tools which provide immediate feedback on the dynamic behavior of a program even while programming" [8]. The use of "live" in this context dates to Tanimoto [6], although systems embodying various levels of liveness date back as far as computing itself.[2]

Why would a programmer want a live programming system? A survey of work in live programming identified 187 papers in the field and identified ten motivations these papers give for liveness [8]. They include improved productivity for programmers (often by speeding up programming tasks), increased accessibility for less experienced programmers, and easier comprehension while reading existing code.

### 2.1 Approaches to Liveness

Developers of live programming systems have experimented widely with the means by which liveness can be achieved. Here, we classify a number of examples into a few imprecise but helpful buckets based on the general approaches taken.

**Liveness outside of code.** Some systems provide quick feedback about the final output of a program, without revealing information on program internals that led to that output. Examples include "live/hot reloading" systems popular for application development [9] and split-screen editors popular for generative art (the P5.js web editor [10], the tree & video-game examples from Victor [5] shown in Figure 1, and TouchDevelop [11]). While the fast feedback these systems provide is certainly an improvement on slow compile-run loops (and often requires sophisticated technical work[3]), this feedback is the most coarsely-grained liveness possible. The examples of liveness we describe in the following categories provide visibility not just into the output of programs, but into their internal operation.

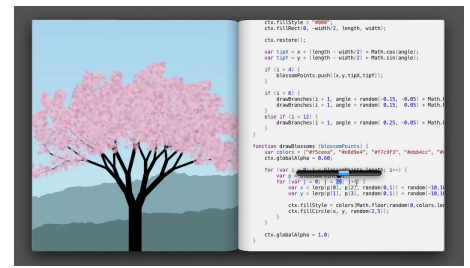

**Figure 1.** A live programming interface shown in Victor [5]. As the code on the right is edited (including with tools like the inline slider shown), the rendered output on the left is immediately updated.

**Liveness within textual code.** One common recipe is to augment textual-code editors with in-context displays showing run-time behavior. Rauch et al.'s "Babylonian-style Programming" [13] surveys the state of the art in this category as of 2012. They evaluate eight existing editors [5], [12], [14]–[18], before adding their own to the list. (Victor [12] is shown in Figure 2.) Notable entrants to this category since 2012 include Swift Playgrounds in Xcode [19] and Projection Boxes [20].

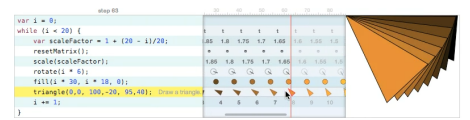

**Figure 2.** A live programming interface shown in Victor [12]. Textual code is entered on the left. A timeline to the right shows the flow of execution, as well as intermediate values and side-effects.

---

2 "Live programming" should be distinguished from "live coding", which has come to refer more narrowly to programming as part of a musical or artistic performance. (See Rein et al. [8] for more on this distinction.)
3 See Burckhardt et al. [11] for an example of technical work in support of liveness.

**Liveness between textual code cells.** Rather than threading live feedback into traditional codebases, other editors provide special interfaces where code is broken up into "cells". These editors can then provide visibility into values flowing between cells. The original instance of this pattern is the spreadsheet. More recently, computational notebooks like Mathematica [21], Jupyter [22], and Observable [23] have extended this model to support more sophisticated computations. Even further afield, Natto reshapes the conventional notebook structure into nodes and wires on a two-dimensional canvas (shown in Figure 3) [24].

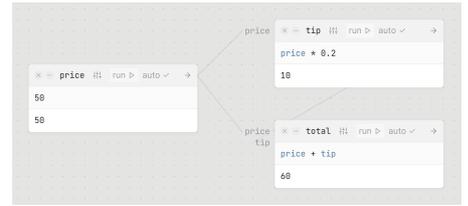

**Figure 3.** A Natto canvas, modified from a built-in example. Each pane runs JavaScript code which can refer to other panes attached with wires, and shows its output value below.

**Liveness within structure editors.** *Structure editors* (sometimes called *projectional editors*) are not based on textual code or cells thereof, but instead allow interactions directly with structured representations of code. Structure editors offer a unique opportunity for live programming. Rather than retrofitting liveness into pre-existing text editors, it can be built into a structure editor from the start. Many projects take advantage of this opportunity, including Subtext [26], Lamdu [25] (shown in Figure 4), Enso [27], and PANE [28].

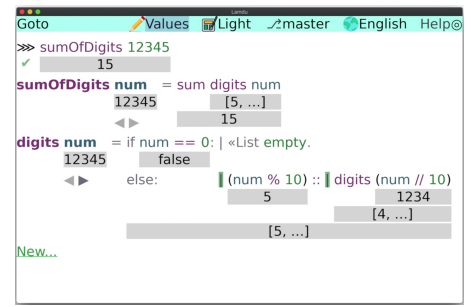

**Figure 4.** The interface of Lamdu. Code is edited in a structured form emulating conventional text. Intermediate values are shown in gray bars under subexpressions. (From Lotem & Chuchem [25].)

**Liveness in a dynamic environment.** Some early live programming environments, including Smalltalk [30], Lisp OSes like Genera [31], and Boxer [29] (shown in Figure 5), do not fit neatly into the above categories. In these environments, the line between code and application is not so clearly drawn. While these environments generally do not provide live visibility into running code, their use often involves executing blocks of code that produce immediate, visible effects.

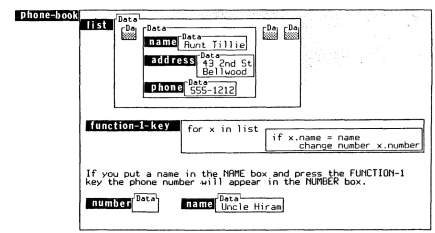

**Figure 5.** A phone book made with Boxer. Procedures, data structures, and interface elements are all represented with nested boxes. (From diSessa & Abelson [29].)

## 3 Richness

People use a vast range of rich, specialized computer interfaces in daily life. We use word processors to write documents, graphics editors to create images, and timeline views for video editing. These interfaces are crafted around their domains, offering representations that afford direct manipulation and visual feedback.

Turning our gaze to static-text programming, this richness disappears. Static text does not afford heterogeneous, domain-specific representations. Parsing input, processing images, and building a visualization are all distinct tasks, but a programmer's screen looks the same no matter which they are working on. Domain-specific languages can offer separate dialects for these separate situations, but even with DSLs, programming still takes on a textual form, deprived of the direct manipulation and visual representations that characterize everyday computer use.

Many efforts from academia and industry have explored bringing this sort of domain-specific richness to programming. Yet there is, to our knowledge, no succinct shared term analogous to "liveness" currently used to describe it. We propose "richness" as a name for this quality. A programming system embodies richness insofar as it allows programmers to edit programs through visualizations and interactions that are tailor-made to the tasks, domains, and contexts they work in. Our test for richness will be whether a system allows programmers to replace writing textual code, at least in some situations, with these custom interactions.[4] Like liveness, richness is meant to be a term of art: just because something feels "rich" in a colloquial sense doesn't mean it will be rich in our specific sense.

Why would a programmer want a rich programming system? We believe many of the motivations for liveness apply here as well, including improving productivity, accessibility, and comprehension of programming systems. But a more specific argument can be made for richness, focused on the way visual representations allow entirely new forms of communication and thought. Andersen et al. [33] write: "[W]riting code means articulating thoughts as precisely as possible… Often these thoughts involve geometrical relationships: tables, nests of objects, graphs, etc. Furthermore, the geometry differs from problem domain to problem domain. To this day, though, programmers articulate their thoughts as linear text." Similarly, Omar et al. [34] write: "Diagrams have played a pivotal role in mathematical thought since antiquity, indeed predating symbolic mathematics. Popular computing and creative tooling, too, has embraced visual representation and direct manipulation interfaces for decades. Programming, however, has remained stubbornly mired in textual user interfaces."

Before digging into examples of rich programming systems, we should clarify what we do not mean by "rich": A programming system being "visual" does not, in and of itself, make it rich in our sense. Many visual programming systems use visual structures to represent the general-purpose, generic structures of textual programming, such as dataflow or control structures. This is not heterogeneous, domain-specific richness.[5] Programming systems that are visual without being rich include block-based programming languages like Scratch [35], node-and-wire editors like Natto [24] and Unreal Engine's "Blueprints" [36], visual structure editors like the "frame-based" editor in Greenfoot [37], and classic "visual programming languages" including VIPR [38], Prograph [39], Forms/3 [40], and Cube [41].

We have picked a particularly narrow scope when drawing the boundaries of "richness", in order to focus on alternatives to static text that capture the rich, heterogeneous interactivity we have come to expect from applications. This definition is still a work in progress. We welcome continued conversation as part of the shared process of sense-making in our community.

### 3.1 Liveness without Richness, Richness without Liveness

Although richness and liveness can be deeply intertwined, they are still distinct qualities. We can make this clear by examining systems that embody liveness without richness or visa versa.

**Liveness without richness.** All the examples we presented in §2 were intentionally chosen to lack richness. This includes, for instance, text-based programming environments like those described in "Babylonian-style Programming" [13]. Though these environments provide visibility into what code is doing, the code itself is still edited as generic static text, without any domain-specific means of program modification.

**Richness without liveness.** In a programming system that is rich but not live, richness often takes the form of "visual syntax" – directly editable visual representations which act as "syntactic sugar" over traditional code. An early example is Erwig & Meyer [42], which shows examples of embedding state diagrams, AVL trees, and pointer-based data structure diagrams into code. Andersen et al. [33] describes a visual macro system for Racket which lets users define interactively editable visual syntaxes

---

[4] By making this restriction, we exclude "read-only richness" (say: domain-specific visualization of data passing through a live system) from our definition of richness. A good example of "read-only richness" is the Glamorous Toolkit [32], a development environment built on top of Pharo Smalltalk offering object inspectors that can be extended with type-specific views. While these inspectors can be visually vibrant and even interactive, programs cannot be constructed or modified through these interactions. Hence, the Glamorous Toolkit provides powerful domain-specific liveness, but not richness.

[5] In support of this distinction, Andersen et al. [33] write: "[visual languages] offer only a fixed set of constructs, though visual ones—meaning a visual language fails to address the problem-specific nature of geometric thought." This echoes Victor [12], who writes: "Traditional visual environments visualize the code. They visualize static structure. But that's not what we need to understand. We need to understand what the code is *doing*. Visualize data, not code. Dynamic behavior, not static structure."

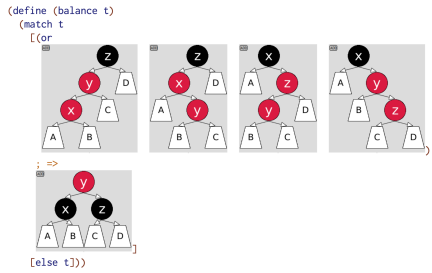
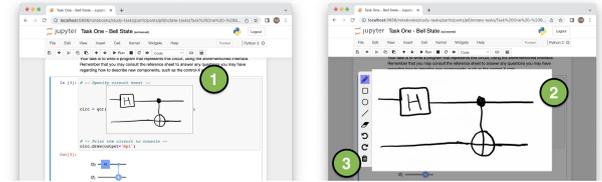

(a) A procedure to balance a red-black tree, implemented using the Racket visual macro system. Here, embedded tree diagrams serve as both patterns to be matched and subexpressions. In both use-cases, they can contain variables. (From Andersen et al. [33].)

(b) The interface to Arawjo et al. [45]'s "notational programming" system. In (1), a canvas is shown embedded inline into code in a Jupyter cell. Selecting this canvas brings up (2), the free-hand editing tool. (From Arawjo et al. [45].)

**Figure 6.** Two examples of richness without liveness.

embedded inside of arbitrary Racket code. They also demonstrate this with tree diagrams (shown in Figure 6a), as well as puzzle games and inline forms. Graphite extends a Java IDE with pop-up "palettes" to bidirectionally edit literals (like colors) in code [43]. MPS (Meta Programming System) is a commercial workbench for developing projectional domain-specific languages. MPS languages typically look mostly like conventional code, but they can embed custom interfaces like tables and state-machine diagrams [44]. Arawjo et al. [45] explores integrating hand-drawn quantum-circuit diagrams into textual code in Jupyter notebooks, an example of what the authors term "notational programming" (shown in Figure 6b). Although all these examples add richness to static text, they remain static – they do not respond to dynamic activity of a running program with live feedback.

### 3.2 Richness with Liveness

We turn now to the intersection of liveness and richness, a combination which opens up synergetic possibilities. Chief among these is the possibility of *programming by demonstration* (PbD) [46]. In a PbD system, a user interacts with concrete objects via direct manipulation. The system is able to generalize from a "demonstration" on one specific input to new inputs, turning the demonstration into a reusable program. Projects have explored how to achieve this goal in different domains, from wrangling tabular data (Wrangler [47]) to building custom visualizations (Drawing Dynamic Visualizations [48], shown in Figure 7a, and Lyra [49]) to building entire web applications (Gneiss [50]). The mechanisms they use vary widely, adapting to the needs of their respective domains. However, all PbD systems must be *live* and *rich*. They must be *live*, because a demonstration is performed on concrete input data, and effects must be shown as the user performs their demonstration. These effects are only available at run-time, so a purely static system cannot support PbD. PbD systems must also be *rich*, because direct manipulation requires domain-specific representations and interactions.

Programming by demonstration is one particularly exciting category of live & rich tools, but it is not the only approach such tools take. Others still require the user to express their intent symbolically, but use liveness and richness to make this process easier. Examples like this include InterState [51] (shown in Figure 7b), Apparatus [52], Object-Oriented Drawing [53], and various regular-expression development tools [54]–[56]. These tools use domain-specific visual syntaxes, like those under the "Richness without liveness" heading of §3.1, but they integrate live feedback into these syntaxes.

## 4 Composability

A pattern is implicit in our previous discussions. When discussing "liveness without richness" and "richness without liveness" (in §3.1), most example systems positioned themselves as general-purpose, roughly the analogue of a conventional general-purpose programming language like Python.

Not so, in our more recent discussion of "richness with liveness" (in §3.2). The tools described in that section were targeted at specific domains. Indeed, their strength came largely from this specificity. Each of these tools replaced generic symbolic programming processes with direct interactions tailored

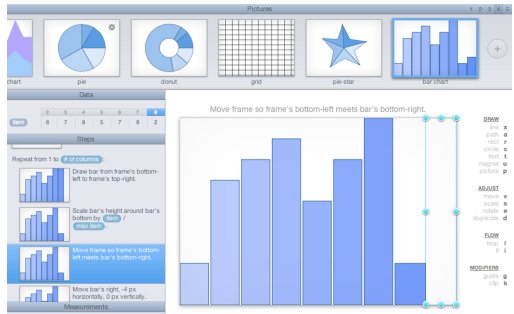 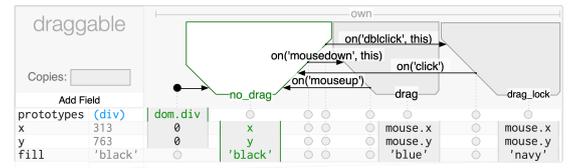

(a) The interface shown in Drawing Dynamic Visualizations. A bar chart has been constructed using a combination of direct manipulations in the canvas view and explicit edits to a script in the left panel. (From Victor [48].)

(b) The editor for an InterState object. States and transitions are shown and edited in the upper panel. Values for properties in different states are edited below, with concrete live values shown on the bottom-left. (From Oney et al. [51].)

**Figure 7.** Two examples of richness with liveness: the first via programming-by-demonstration, the second not.

specifically to their domain.

However, this leaves an important question on the table: How can these powerful but specialized tools be combined together to solve real-world programming goals? Real-world programming often requires performing multiple tasks in concert. For instance, suppose a data analyst wants to understand how well a product launch is proceeding. They may need to: 1. acquire data from an API, 2. reformat this data from semi-structured JSON into a tabular format, 3. wrangle the tabular data further, and 4. build charts to show patterns in this data. Each of these steps could be supported by a targeted live and rich tool. For instance, parts of Gneiss [50] could support steps 1 & 2, Wrangler [47] could support step 3, and Lyra [49] could support step 4. In what sort of environment could the analyst fluidly make use of these tools *together*?

We are seeking *composability*: the ability to freely combine smaller programmed artifacts into larger ones, to accomplish larger goals. Composability is a pervasive quality of conventional programming, visible whenever we write Python scripts that call out to multiple libraries, or even simply when we nest `if` and `for` inside one another. Ultimately, the ability to freely combine building blocks is what gives programming its unlimited expressive potential.

The live-but-not-rich and rich-but-not-live systems we looked at in §3.1 followed the model of conventional programming and maintained composability. But, tellingly, the live & rich systems discussed in §3.2 were implemented as standalone applications, rather than as composable components in some larger programming system. This trend shows the difficulties of achieving composability when liveness and richness are combined. Neither operating-system-level features nor programming platforms (until recently) allow for live composition of rich interfaces. Unfortunately, this makes live & rich tools harder to use in real-world situations that require combining multiple tools, such as the one faced by the data analyst described above.

The main question this paper raises is: How might programming acquire the desirable qualities of *liveness* and *richness* without giving up the crucial quality of *composability*? In this section, we first discuss a few current approaches which we believe do *not* provide full composability, before surveying two recent systems which successfully combine all three qualities from the start.

## 4.1 Current Approach: Manual Integration

The live, rich, domain-specific tools we discussed in §3.2 usually provide *some* forms of integration with the outside world, if only for demonstration. This may take the form of copying and pasting text snippets, like copying regular expressions in and out of a regular-expression development tool, or copying generated code out of (and back into) Wrangler [47]. Alternatively, it may take the form of importing and exporting files, like how tabular data is loaded into and out of Wrangler. In any case, we call this *manual* integration, as it requires the user propagating changes between the tool and its surroundings by hand. On some level, manual integration works. A certain workflow is possible, copying program snippets and data back and forth between a tool and a codebase. However, this workflow leaves much to be desired.

Most importantly, there is a lack of *overall liveness*. If a situation requires the use of multiple composed tools, work in one tool may produce downstream effects that affect another tool. For instance, if the data analyst described above switches to getting data from a new API, this may affect the way they need to wrangle their data. For their overall workflow to be live, in a holistic sense, the effects of their changes should be visible as quickly as possible in the context of downstream tools, with as little distracting effort as possible. However, in a manually-integrated system, the programmer must propagate changes to a tool's inputs and outputs by hand. Although their tools may be live *internally*, there is no liveness across the boundaries between tools.

Manual integration carries other costs. There is a lack of *collocation*. Using a separate tool requires leaving a programmer's preferred environment and entering a new space. With this comes a potential for disorientation and distraction. Furthermore, there is lack of *full persistence*. For instance, although a programmer may curate a set of test inputs to a regular-expression helper tool as they use it, this ensemble is not persisted when the regular expression is copy-pasted back to code.

These problems add up to a significant cost when integrating tools manually. Because this cost occurs on a per-tool basis, it disincentivizes the development of small tools, even though, integration cost aside, ensembles of small tools would allow for greater flexibility and combinatorial possibility. In an environment with high integration costs, only large, monolithic tools will survive.

## 4.2 Current Approach: Bundles

If manual integration carries heavy costs, a natural alternative is to bundle together, in a single tool, multiple live & rich tools a user would want to use for a particular larger-scale task.

A good example of this approach can be found in Gneiss [50], shown in Figure 8. Gneiss is a tool for making interactive, data-driven web applications. It combines, in three panes, three separate tools: an API query tool, a spreadsheet, and an interface builder. These tools support a fluid workflow where the user crafts an API call, drags pieces of the response into the spreadsheet, and builds an interface off of this spreadsheet. Data flows between these tools automatically, allowing an integrated live experience that would not be possible with tools that were manually integrated into a static codebase. Gneiss makes a compelling demonstration of the benefits of integrated liveness – liveness not just within tools, but between them.

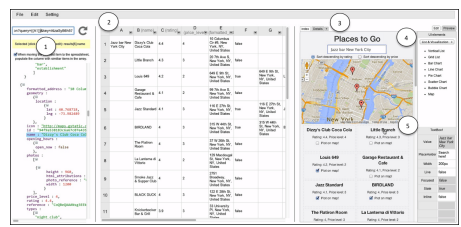

**Figure 8.** The Gneiss interface, consisting of three panes: (1) is where a call to an API is crafted and results are shown. These results can be dragged into (2), a spreadsheet interface. (3) is a direct-manipulation web interface builder, driven by data in the spreadsheet. (From Chang & Myers [50].)

However, Gneiss makes this integrated liveness available only for one particular, hard-wired bundle of tools. This means that, if a user's workflow is close to, but not quite, the workflow anticipated by Gneiss's creators, Gneiss will fail to meet their needs. What if a user wants to build an interface off of imported CSV data, rather than JSON from an API? What if a user wants to construct a bespoke chart from their data, rather than a web app? Even if tools exist to support these steps, a Gneiss user cannot incorporate them.

A similar limitation holds in the other direction. To build Gneiss, its creators put significant work into designing and implementing three ingenious live tools. The domains of these tools presumably extend further than the specific workflow that Gneiss was designed around. Since these tools are built into Gneiss, however, they cannot be brought into new environments or contexts.

If we judge Gneiss by the standards of modern application software, these limitations are not surprising. We have not come to expect open-ended reconfigurability and modularity in the world of graphical software interfaces.[6] From the perspective of programming systems, however, these limitations are striking. If Gneiss's capabilities were represented with traditional code, they would look like a fixed script, making use of three specific libraries connected in one particular way. A user's

---

6 A partial exception to this is applications supporting plugin or extension mechanisms, such as web browsers and platform-like creative tools.

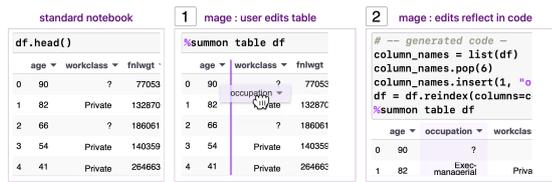
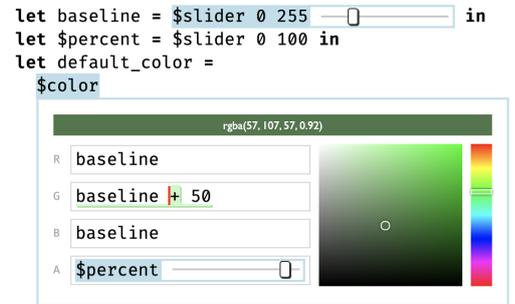

(a) The mage workflow: A live tool is summoned from a notebook cell. Direct-manipulation edits in the live tool are written back, as code, to the cell they were summoned from. (From Kery et al. [59].)

(b) An example use of Livelits in Hazel. The color picker and various sliders are Livelits, embedded into Hazel code (and each other). (From Omar et al. [34].)

**Figure 9.** Two systems combining liveness, richness, and composability.

interactions with Gneiss would modify blank slots within these library calls. However, the user cannot import additional libraries, or rearrange the existing libraries into a new structure. The open-ended composability that is present in conventional programming is absent in the world of graphical interfaces, even live and rich interfaces like Gneiss.

### 4.3 Current Approach: Non-persistent Tools

At first glance, Observable [23] may seem to embody liveness and richness together with composability. As a cell-based computational notebook, it certainly embodies liveness. Furthermore, Observable cells can render arbitrary HTML as part of their output, including interactive widgets, seemingly embodying richness. Because interactions in these widgets can affect the output of the cell, values can flow from the output of one interactive widget to the input of another. This would seem to enable composition of live, rich tools.

Although Observable is indeed close to this goal, it fails to achieve it. The most important limitation standing in the way is that, although Observable cells can render UI, this UI cannot persistently modify an underlying program. That is, even though an Observable cell can render an arbitrarily rich tool as part of its output, interactions with this tool cannot be "saved" back to the notebook, and their effects will always disappear when the notebook is reloaded. If programming tools are to be part of the process of *writing* programs, rather than just ephemeral use thereof, rich UIs must be able to modify the underlying program.

The appeal of live, rich (& persistent) tools is strong enough that several platforms, including Observable, have added them as one-off features. Observable has a "Data Table Cell" feature, which lets the user embed a live, rich direct-manipulation table editor in place of an ordinary cell [57]. Similarly, the computational notebook Hex offers several "no-code" cells: live, rich tools for building charts, pivoting and filtering data, and other specific tasks [58]. In both cases, the tools are specially built into their respective platforms and can access persistence features that are not available to third-party tools.

### 4.4 Liveness, Richness, and Composability

We have seen that attempts to compose tools together manually fail to preserve liveness across tool boundaries, and attempts to compose tools into pre-made bundles fail to provide users with open-ended composability. For a system to overcome these limitations, it needs to provide open-ended composition of persistent, rich interfaces into an integrated live system. Two recent projects have found their way into this triple-intersection of liveness, richness, and composability.

**mage.** mage is "an API for allowing smooth transitions between GUI and code work in notebooks", implemented as a Jupyter Notebook extension [59]. Tools made with mage can be launched from cells of Jupyter Notebooks. A tool is rendered with its cell, and it can read from and write to the code in that cell. This workflow is depicted in Figure 9a. mage is demonstrated with a variety of tools, including a chart builder, an image editor, and a table editor which converts direct-manipulation actions into dataframe manipulation code.

mage is *live*: Interacting with a tool causes its cell's code to be rewritten and rerun automatically. Cells with tools can be quickly re-run to reflect any changes in their inputs. mage is *rich*: Tools implemented using mage's API can make use of all the features available on the web platform to offer domain-specific visualizations and interactions. A tool can write code back to its Jupyter cell host, making it a persistent programming tool rather than just an ephemeral run-time interface. mage is *composable*: This composability comes from mage's embedding into a Jupyter notebook. Multiple mage tools can be used in concert by placing them into separate cells, feeding data between them with variables.

**Livelits.** Livelits (live literals) are an extension to Hazel [60], "a live functional programming environment designed around hole-driven development" [34]. Hazel itself is a live structured editor which emulates conventional statically-typed functional code. Livelits add "user-defined GUIs embedded persistently into code" [34] to this. Livelits are demonstrated with a few such GUIs, including sliders, color pickers, table literals, and image editors. A few of these are shown working together in Figure 9b. While mage tools exist at the level of Jupyter cells, allowing only a "flat" structure of embedding, livelits exist at the level of Hazel AST nodes. This means that livelits and textual Hazel code can be arbitrarily nested within each other, recursively.

Livelits are *live*: This liveness ultimately comes from Hazel, as mage's liveness comes from Jupyter. Hazel's liveness is deeper than Jupyter's in several ways. In Hazel, live feedback is available on the values of arbitrary subexpressions, not just on the cell level. Furthermore, Hazel runs reactively, automatically propagating changes without the user explicitly re-running cells. Livelits are *rich*: Like mage tools, they have access to the web platform, allowing diverse and vibrant interfaces. They persist their state (an arbitrary serializable model) directly into a program's AST, rather than generating and re-parsing textual code as mage tools do. Livelits are *composable*: They can occur anywhere a Hazel expression can occur. Not only can livelits be chained together through variable assignments (analogous to composition of mage tools in Jupyter cells), but they can also be embedded directly into one another, into function definitions, etc.

**Discussion.** mage and livelits enable something we believe was not previously possible: the integration of domain-specific interactive programming tools into general-purpose programming environments, with full liveness available to support techniques like programming by demonstration. Imagine the live & rich tools we surveyed in §3.2, from visualization editors to direct-manipulation data wranglers, rebuilt as tools in mage or in livelits. Built this way, they could be incorporated into more powerful workflows, combined with other tools or with conventional textual code. mage, livelits, or a platform like them could unlock new value for live & rich tools by enabling composition.

mage and livelits have contrasting limitations standing in the way of this vision. mage tools are flexible and powerful, but because they are summoned at the level of Jupyter cells, they cannot be embedded into general programming constructs like loops or function definitions. Furthermore, while Jupyter is directly useful in contexts like data work, it is not immediately clear how to generalize mage's Jupyter-based approach to other contexts like application development. On the other hand, livelits exist as a first-class language construct and can be embedded anywhere in the Hazel language. However, Hazel's strict typing discipline limits their flexibility. Many of the tools we have discussed, such as Gneiss's sub-tools and Wrangler, cannot exist as livelits because their behavior cannot be expressed in Hazel's type system. As the name "live literals" suggests, livelits have so far been restricted to simple widgets, like sliders and color pickers, for constructing simple values.

Currently, neither mage nor livelits are available for public use. We wonder whether a publicly available platform for composition could facilitate research into experimental live & rich tools. Suppose a researcher has an idea for a novel live & rich tool. As effective as their tool may be for its task, evaluating the tool requires building enough complementary infrastructure that the tool can be demonstrated in an end-to-end workflow. This may involve a significant amount of work which is unrelated to the novel interactions they are exploring. If the researcher could implement their tool in a system comparable to mage or livelits, they could test it in the context of complementary tools and conventional-code escape hatches. In this way, a publicly available platform for composition could lower the barrier to entry for live & rich tool research and accelerate innovation in this field.

## 5 Conclusion

Liveness and richness present complementary directions for efforts beyond static-text programming. Liveness gives programmers visibility into programs' internal behavior, reacting to their edits, and richness gives programmers the ability to edit programs via visual representations that match the domains they work with. These directions can be pursued separately, but their combination presents even greater possibilities, enabling new techniques like programming by demonstration.

Although live & rich programming interfaces hold much potential, their usefulness in practice is limited when they are implemented as stand-alone applications. Prevailing approaches like manual integration and bundles fail to replicate the free-form composability of conventional code, at least not without losing overall liveness in the process.

mage and livelits address this lack of composability. They follow a shared pattern, extending existing general-purpose programming environments (Jupyter and Hazel, respectively) with live & rich tools. Could other live programming environments, from text-based environments discussed in "Babylonian-style Programming" [13] to notebooks like Observable [23] and visual editors like Natto [24], be similarly extended? Or, to lay out an even more ambitious goal, could tools be shared between these environments, so that a diversity of environments could be explored alongside a diversity of tools?

mage and livelits have only begun to explore the design space at the triple intersection of liveness, richness, and composability. We hope, above all else, that by articulating this combination of qualities we can encourage further ventures into this emerging space.

## Acknowledgement

We thank Geoffrey Litt, Paula Te, the UW Interactive Data Lab, and the PLATEAU workshop community for their valuable feedback. This work was supported by a Moore Foundation software grant.

## References

[1] I. E. Sutherland, "Sketchpad," in *Proceedings of the May 21-23, 1963, spring joint computer conference on - AFIPS '63 (Spring)*, ACM Press, 1963.

[2] T. O. Ellis, J. F. Heafner, and W. L. Sibley, "The GRAIL PROJECT: An EXPERIMENT IN MAN-MACHINE COMMUNICATIONS," 1969.

[3] T. Ellis, J. Heafner, and W. Sibley, *The GRAIL Language and Operations*. RAND Corporation, 1969.

[4] B. A. Nardi, *A small matter of programming* (A Small Matter of Programming). London, England: MIT Press, Jul. 1993.

[5] B. Victor, "Inventing on Principle," https://vimeo.com/36579366, Presented at the the Canadian University Software Engineering Conference (CUSEC), 2012.

[6] S. L. Tanimoto, "Viva: A visual language for image processing," *Journal of Visual Languages and Computing*, vol. 1, no. 2, pp. 127–139, Jun. 1990.

[7] *Live 2013: Workshop on Live Programming*.

[8] P. Rein, S. Ramson, J. Lincke, R. Hirschfeld, and T. Pape, "Exploratory and Live, Programming and Coding," *The Art, Science, and Engineering of Programming*, vol. 3, no. 1, Jul. 2018.

[9] D. Abramov, *Live React: Hot Reloading with Time Travel*, Jul. 2015.

[10] p5.js, *P5.js Web Editor*.

[11] S. Burckhardt, M. Fahndrich, P. de Halleux, *et al.*, "It's alive! continuous feedback in UI programming," *ACM SIGPLAN Notices*, vol. 48, no. 6, pp. 95–104, Jun. 2013.

[12] B. Victor, *Learnable Programming*, http://worrydream.com/LearnableProgramming/, 2012.

[13] D. Rauch, P. Rein, S. Ramson, J. Lincke, and R. Hirschfeld, "Babylonian-style Programming: Design and Implementation of an Integration of Live Examples into General-purpose Source Code," *The Art, Science, and Engineering of Programming*, vol. 3, no. 3, Feb. 2019.

[14] J. Edwards, "Example centric programming," *ACM SIGPLAN Notices*, vol. 39, no. 12, pp. 84–91, Dec. 2004.

[15] C. Granger, *Light Table*, http://lighttable.com/, 2022.

[16] T. Imai, H. Masuhara, and T. Aotani, "Shiranui: A live programming with support for unit testing," in *Companion Proceedings of the 2015 ACM SIGPLAN International Conference on Systems, Programming, Languages and Applications: Software for Humanity*, ACM, Oct. 2015.

[17] S. Kasibatla and A. Warth, "Seymour: Live Programming for the Classroom," https://harc.github.io/seymour-live2017/, Presented at the Workshop on Live Programming (LIVE) 2017, 2017.

[18] T. van der Storm and F. Hermans, "Live Literals," https://homepages.cwi.nl/~storm/livelit/livelit.html, Presented at the Workshop on Live Programming (LIVE) 2016, 2016.

[19] Apple Inc., *Wwdc 2014*, https://www.youtube.com/watch?v=w87fOAG8fjk, 2014.

[20] S. Lerner, "Projection Boxes: On-the-fly Reconfigurable Visualization for Live Programming," in *Proceedings of the 2020 CHI Conference on Human Factors in Computing Systems*, ACM, Apr. 2020.

[21] S. Wolfram, *The Mathematica Book*, 5th edition. Champaign, Ill.: Wolfram Media Inc, Aug. 2003.

[22] T. Kluyver, B. Ragan-Kelley, F. Pérez, et al., "Jupyter Notebooks - a publishing format for reproducible computational workflows," in *International Conference on Electronic Publishing*, 2016.

[23] Observable Inc., *Observable - Explore, analyze, and explain data. As a team.* https://observablehq.com/, 2022.

[24] P. Shen, *Welcome! – natto*, https://natto.dev/, 2022.

[25] E. Lotem and Y. Chuchem, *Lamdu*, https://www.lamdu.org/, 2022.

[26] J. Edwards, "Subtext," in *Proceedings of the 20th annual ACM SIGPLAN conference on Object-oriented programming, systems, languages, and applications*, ACM, Oct. 2005.

[27] Enso, *Enso*, https://enso.org/, 2022.

[28] J. Horowitz, "Pane: Programming with Visible Data," http://joshuahhh.com/projects/pane/, Presented at the Workshop on Live Programming (LIVE) 2018, 2018.

[29] A. A. diSessa and H. Abelson, "Boxer: A reconstructible computational medium," *Communications of the ACM*, vol. 29, no. 9, pp. 859–868, Sep. 1986.

[30] A. Goldberg and D. Robson, *Smalltalk-80: the language and its implementation*. Addison-Wesley Longman Publishing Co., Inc., 1983.

[31] J. Walker, D. Moon, D. Weinreb, and M. McMahon, "The Symbolics Genera Programming Environment," *IEEE Software*, vol. 4, no. 6, pp. 36–45, Nov. 1987.

[32] T. Girba, *Glamorous Toolkit*.

[33] L. Andersen, M. Ballantyne, and M. Felleisen, "Adding interactive visual syntax to textual code," *Proceedings of the ACM on Programming Languages*, vol. 4, pp. 1–28, Nov. 2020.

[34] C. Omar, D. Moon, A. Blinn, I. Voysey, N. Collins, and R. Chugh, "Filling typed holes with live GUIs," in *Proceedings of the 42nd ACM SIGPLAN International Conference on Programming Language Design and Implementation*, ACM, Jun. 2021.

[35] M. Resnick, J. Maloney, A. Monroy-Hernández, et al., "Scratch," *Communications of the ACM*, vol. 52, no. 11, pp. 60–67, Nov. 2009.

[36] I. Epic Games, *Blueprints Visual Scripting*.

[37] M. Kölling, N. C. C. Brown, and A. Altadmri, "Frame-Based Editing," in *Proceedings of the Workshop in Primary and Secondary Computing Education*, ACM, Nov. 2015.

[38] W. Citrin, M. Doherty, and B. G. Zorn, "The Design of a Completely Visual Object-Oriented Programming Language," 1994.

[39] P. Cox, F. Giles, and T. Pietrzykowski, "Prograph: A step towards liberating programming from textual conditioning," in *[Proceedings] 1989 IEEE Workshop on Visual Languages*, IEEE Comput. Soc. Press.

[40] M. M. Burnett and A. L. Ambler, "Interactive Visual Data Abstraction in a Declarative Visual Programming Language," *Journal of Visual Languages amp; Computing*, vol. 5, no. 1, pp. 29–60, Mar. 1994.

[41] M. Najork and S. Kaplan, "The CUBE languages," in *Proceedings 1991 IEEE Workshop on Visual Languages*, IEEE Comput. Soc. Press.

[42] M. Erwig and B. Meyer, "Heterogeneous visual languages-integrating visual and textual programming," in *Proceedings of Symposium on Visual Languages*, IEEE Comput. Soc. Press.

[43] C. Omar, Y. S. Yoon, T. D. LaToza, and B. A. Myers, "Active code completion," in *2012 34th International Conference on Software Engineering (ICSE)*, IEEE, Jun. 2012.

[44] JetBrains s.r.o., *Mps: The Domain-Specific Language Creator by JetBrains*, https://www.jetbrains.com/mps/, 2022.

[45] I. Arawjo, A. DeArmas, M. Roberts, S. Basu, and T. Parikh, "Notational Programming for Notebook Environments: A Case Study with Quantum Circuits," in *The 35th Annual ACM Symposium on User Interface Software and Technology*, ACM, Oct. 2022.

[46] A. Cypher, Ed., *Watch what I do: Programming by demonstration*. The MIT Press, 1994.

[47] S. Kandel, A. Paepcke, J. Hellerstein, and J. Heer, "Wrangler," in *Proceedings of the SIGCHI Conference on Human Factors in Computing Systems*, ACM, May 2011.

[48] B. Victor, "Drawing Dynamic Visualizations," https://vimeo.com/66085662, Presented at the Stanford HCI seminar on February 1, 2013, 2013.

[49] A. Satyanarayan and J. Heer, "Lyra: An Interactive Visualization Design Environment," *Computer Graphics Forum*, vol. 33, no. 3, pp. 351–360, Jun. 2014.

[50] K. S.-P. Chang and B. A. Myers, "Creating interactive web data applications with spreadsheets," in *Proceedings of the 27th annual ACM symposium on User interface software and technology*, ACM, Oct. 2014.

[51] S. Oney, B. Myers, and J. Brandt, "Interstate," in *Proceedings of the 27th annual ACM symposium on User interface software and technology*, ACM, Oct. 2014.

[52] T. Schachman and J. Horowitz, *Apparatus: A Hybrid Graphics Editor and Programming Environment for Creating Interactive Diagrams*.

[53] H. Xia, B. Araujo, T. Grossman, and D. Wigdor, "Object-Oriented Drawing," in *Proceedings of the 2016 CHI Conference on Human Factors in Computing Systems*, ACM, May 2016.

[54] F. Dib, *Regex101: Build, Test, and Debug Regex*.

[55] S. Toarca and P. Prelich, *Debuggex: Online Visual Regex Tester. JavaScript, Python, and PCRE*.

[56] gskinner, *Regexr: Learn, Build, Test RegEx*.

[57] Observable Inc., *Quickly Explore Analyze Your Data With Data Table Cell / Observable / Observable*, https://observablehq.com/@observablehq/introducing-data-table-cell, 2022.

[58] I. Hex Technologies, *Hex - Do More with Data, Together*.

[59] M. B. Kery, D. Ren, F. Hohman, D. Moritz, K. Wongsuphasawat, and K. Patel, "Mage: Fluid Moves Between Code and Graphical Work in Computational Notebooks," in *Proceedings of the 33rd Annual ACM Symposium on User Interface Software and Technology*, ACM, Oct. 2020.

[60] C. Omar, I. Voysey, R. Chugh, and M. A. Hammer, "Live functional programming with typed holes," *Proceedings of the ACM on Programming Languages*, vol. 3, pp. 1–32, Jan. 2019.